\documentclass{jaa}
\usepackage{txfonts}
\usepackage{graphicx}
\usepackage{color}

\usepackage[authoryear]{natbib}
\newcommand{\farcs}{\mbox{\ensuremath{.\!\!^{\prime\prime}}}}
\newcommand{\farcm}{\mbox{\ensuremath{.\!\!^{\prime}}}}
\begin{document}
\title[Velocity Fields and Galaxy Interaction in the Quartet of Galaxies]{H$\alpha$ Velocity Fields and Galaxy Interaction in the Quartet of Galaxies NGC 7769, 7770, 7771 and 7771A}
\author[A.~A.~Yeghiazaryan, T.~A.~Nazaryan \& A.~A.~Hakobyan]%
       {A.~A.~Yeghiazaryan, T.~A.~Nazaryan\thanks{\emph{e-mail:}~nazaryan@bao.sci.am} \& A.~A.~Hakobyan\thanks{\emph{e-mail:}~hakobyan@bao.sci.am}\\
       Byurakan Astrophysical Observatory, 0213 Byurakan, Aragatsotn province, Armenia.}
\maketitle
\label{firstpage}
\begin{abstract}
The quartet of galaxies NGC~7769, 7770, 7771 and 7771A is
a system of interacting galaxies.
Close interaction between galaxies caused
characteristic morphological features:
tidal arms and bars, as well as an induced star formation.
In this study, we performed the Fabry-Perot
scanning interferometry of the system in H$\alpha$ line
and studied the velocity fields of the galaxies.
We found that the rotation curve of NGC~7769 is weakly distorted.
The rotation curve of NGC~7771 is strongly
distorted with the tidal arms caused by
direct flyby of NGC~7769 and flyby of a smaller neighbor NGC~7770.
The rotation curve of NGC~7770 is significantly skewed because of
the interaction with the much massive NGC~7771.
The rotation curves and morphological disturbances
suggest that the NGC~7769 and NGC~7771
have passed the first pericenter stage,
however, probably the second encounter has not happened yet.
Profiles of surface brightness of NGC 7769 have a characteristic break,
and profiles of color indices have a minimum
at a radius of intensive star formation induced by the interaction with NGC~7771.
\end{abstract}

\begin{keywords}
Galaxies: interactions---galaxies: star formation---galaxies: peculiar---galaxies:
individual: NGC~7769, NGC~7770, NGC~7771, NGC~7771A---supernovae: individual: 2003hg.
\end{keywords}

\section{Introduction}

More than 30 years ago, NGC~7769, 7770 and 7771 galaxies, among the few hundreds of others,
were included in the list of galaxies with ultraviolet (UV) excess with numbers
Kaz~346, 347, and 348 respectively (\citealt{1980Ap.....16....7K,2010Ap.....53...57K}).
Direct observations of these galaxies
at the primary focus of 2.6-m telescope of the Byurakan Astrophysical Observatory (BAO, Armenia)
have shown that they are all spirals (\citealt{1983Ap.....19..345E}).
Spectral observations of galaxies NGC~7769, 7770 and 7771,
conducted with 6-m BTA telescope of the Special Astrophysical Observatory (Russia),
have shown that forbidden lines of sulfur and oxygen,
and Balmer emission lines of hydrogen are present in their spectra (\citealt{1989Ap.....30..355K}).
It was found that galaxies NGC~7769, 7770 and 7771 together are
a physical triplet of galaxies located at a distance of $60\pm4$~Mpc.
Later, these galaxies have been studied
as an isolated quartet including also the small galaxy NGC~7771A (\citealt{1997AJ....114...77N}).
The system was also observed in 21-cm HI line (\citealt{1997AJ....114...77N,2005A&A...442..137N}).

In the current work, we report results of the optical interferometry of the system
and analyze the kinematics of galaxies. A detailed description of the morphological
features of the galaxies is presented. We also discuss the influence of interaction on
the kinematics, dynamics and star formation in the system.
Known models of galaxy interactions are based mostly on statistical observational data.
We try to illustrate how and to what extend these models can be applied to
explain the features of the galaxies in this system.

\begin{figure}
  \label{syste}
  \centering
  \includegraphics[width=0.8\hsize]{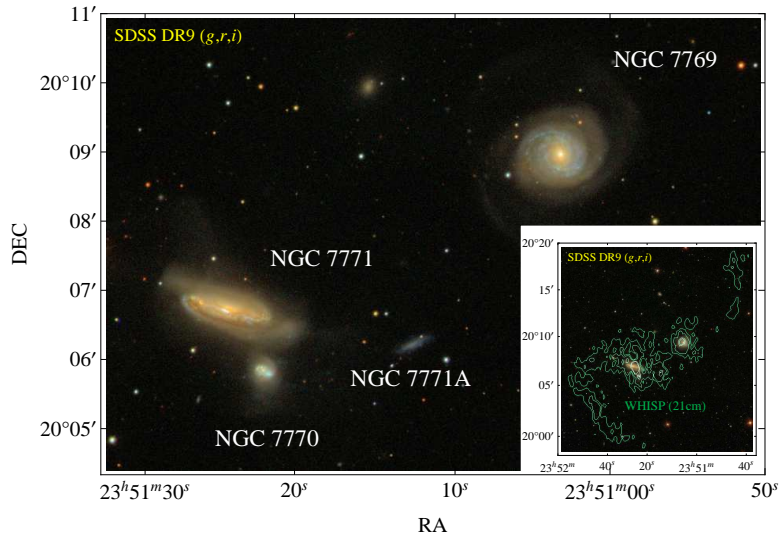}\\
  \parbox{1.0\hsize}{\caption{\small{The SDSS \emph{g, r, i} image of the system.
         Inset: a broader region containing the system, where
         tidal arms extend much further in the HI radio waveband.
         Intensity contours of the WHISP 21-cm radio map (\citealt{2005A&A...442..137N})
         are overlapped on the SDSS image.}}}
\end{figure}

\section{Description of the system}

Using \emph{gri} color images of the Sloan Digital Sky Survey
(SDSS; \citealt{2014ApJS..211...17A}),
we morphologically classified the galaxies of the system,
measured their geometric and photometric parameters,
as well as analyzed color profiles of NGC~7769.
Table~\ref{tab} presents parameters of the galaxies.
Figure~\ref{syste} shows SDSS \emph{g, r, i} image of the system.
Below we pay attention to some remarkable properties of the galaxies of the system
and analyze them in order to illustrate some interesting features of processes
caused by close interaction of the galaxies.

\begin{table}
	\centering
	\caption{Parameters of galaxies in the quartet.}
	\label{tab}
	\tabcolsep 4pt
	\begin{tabular}{lrrrr}
	\hline
	NGC & 7769 & 7770 & 7771 & 7771A \\
	Kaz & 346 & 347 & 348 & -- \\
	\hline
    RA~(deg) & 357.76659 & 357.84442 & 357.85435 & 357.80505 \\
	Dec~(deg) & 20.15045 & 20.09680 & 20.11160 & 20.103411 \\
	$V_{\rm r}$~${\rm (km~s^{-1})}$ & 4220 & 4045 & 4260 & 4060 \\
	Morph. & Sb & SABc & SBb & Sm\\
	Arm~class & 7 & 4 & 12 & -- \\
	$g-{\rm band}~D_{25}$ (arcsec) & 196 & 65 & 250 & 50 \\
	PA & $30^\circ$ & $60^\circ$ & $70^\circ$ & $114^\circ$ \\
	incl. & $40^\circ$ & $45^\circ$ & $65^\circ$ & $70^\circ$ \\
	$g$ mag & $12.61\pm0.04$ & $14.16\pm0.02$ & $12.47\pm0.01$ & $16.70\pm0.07$ \\
	$u-g$ & $1.35\pm0.11$  & $1.20\pm0.05$ & $1.51\pm0.03$ & $1.14\pm0.17$ \\
	$g-r$ & $0.75\pm0.07$ & $0.69\pm0.04$ & $0.86\pm0.02$ & $0.50\pm0.12$ \\
	$g-i$ & $1.14\pm0.07$ & $1.01\pm0.03$ & $1.32\pm0.01$ & $0.70\pm0.12$ \\
	$g-z$ & $1.63\pm0.06$ & $1.33\pm0.03$ & $1.73\pm0.01$ & $1.14\pm0.01$ \\
	$\log{(m /M_\odot)}$ & $10.27\pm0.15$ & $9.56\pm0.11$ & $10.49\pm0.09$ & $8.26\pm0.20$\\
  \hline \\
  \end{tabular}
\end{table}

The structure of spiral arms of NGC~7769 can be classified
as having an arm-class 7 (grand-design) according to the scheme in
\citet{1987ApJ...314....3E}: two symmetric long outer arms and flocculent inner arms.
Spiral arms of NGC~7771 have arm-class 12 (grand-design)
according to the scheme in \citet{1987ApJ...314....3E}: two symmetric long arms dominating in the disk.
Radio observations in \citet{1997AJ....114...77N} and
\citet{2005A&A...442..137N} showed that the southern arm of NGC~7771
extends much further than what is visible in the optical image,
making a bridge to NGC~7771A to the West and ending with a region of bright radio emission,
see Figure~\ref{syste}.
The northern arm extends up to about 10-15 optical radii of the galaxy (16~arcmin)
and becomes an arc stretched around NGC~7770 to the South.
The long bar (21~kpc) of NGC~7771 also stands out.
The spiral arms of NGC~7771, as well as outer arms of NGC~7769
have tidal origin (e.g. \citealt{1972ApJ...178..623T,1979ApJ...233..539K,2011MNRAS.414..538K}).
The existence of bar is a characteristic feature of most grand-design galaxies
(e.g. \citealt{1989ApJ...342..677E}).
Probably, the bar of NGC~7771 also has a tidal origin (see, \citealt{1987MNRAS.228..635N}).
The system of NGC~7771, 7770 and 7771A is embedded in a
common envelope of neutral hydrogen (\citealt{2005A&A...442..137N}).
Although NGC~7771A has not been included in the list of Kazarian galaxies
(obviously, because of low surface brightness), however,
it is the bluest object in the quartet (see colors in Table~\ref{tab}),
which evidently suggests that it also has a UV excess
caused by a high rate of star formation.

On the SDSS optical images toward the North and South from NGC~7771,
there are some visible faint regions of diffuse emission
probably associated with stars and hot gas
driven out by the minor interaction with NGC~7770 (\citealt{2012MNRAS.425L..46A}).
A comparison of the SDSS images of different colors shows that
NGC~7770 has a weak but long enough bar (from NE to SW).
Spiral arms of NGC~7770 have arm-class 4 according to scheme in \citet{1987ApJ...314....3E}:
one remarkable distorted arm, and the second arm is much shorter.
Because of the low resolution of images, however,
in earlier works of \citet{1983Ap.....19..345E} and
\citet{1997AJ....114...77N}, it was concluded that NGC~7770 has two nuclei.
However photometric analysis of the SDSS images of five colors shows
that NGC~7770 has seven bright HII regions with intensive star formation,
some of which have brightness comparable with the brightness of the nucleus of the galaxy.
Both intensive star formation and significant disturbances of arms
are caused by the interaction with more massive (about 8 times) neighbor
NGC~7771 (e.g. \citealt{2007A&A...468...61D}).
In \citet{2012MNRAS.425L..46A}, it was concluded that NGC~7770 has already experienced
first passage around NGC~7771.

Supernova (SN) 2003hg was discovered in NGC~7771 galaxy
at the coordinates ${\rm RA~(deg)} = 357.85054$, ${\rm Dec~(deg)} = 20.11064$,
and was classified as Type~II SN (collapse of a massive stellar core),
discovered immediately after the explosion (\citealt{2003IAUC.8187....2E}).
Reddening in the spectrum of the SN caused by dust in the host galaxy is significant.
However, direct inspection of the explosion site on the
SDSS \emph{u}- and \emph{g}-band images allows us to see an overlap of the SN
with a giant region of intensive star formation located
at the West side of galaxy bar.
Such a location of SN is in good agreement with
the known correlations of core-collapse (Types Ibc and II) SNe
with star formation in galaxies (e.g. \citealt{2006A&A...453...57J,2008A&A...488..523H}),
including also star formation induced by an interaction with
neighbor galaxies (e.g. \citealt{2013Ap&SS.347..365N,2014MNRAS.444.2428H}).

\section{Observations}

In order to study the velocity fields of the galaxies,
the observations were carried out at the 2.6-m telescope of BAO
on 8 November 1996, with the ByuFOSC
(Byurakan Faint Object Spectral Camera) in the interferometric
mode, attached at the prime focus of the telescope. This device,
as well as the pointing-guiding system `Bonnette', were
designed and assembled at the Marseille Observatory in 1996.
ByuFOSC includes a focal reducer (bringing the original F/4 focal
ratio of the prime focus to F/2) with parallel beam allowing the
installation of the various dispersing elements, filter wheel, and
a CCD detector (Thomson $1028 \times 1060$ matrix with ${\rm 5e^{-}}$ rms
read-out noise and 19~${\rm \mu}$m pixel size).
The scanning Fabry-Perot interferometer
was placed in the parallel beam. The detector was used
in half-obscured mode, allowing a quick shift of an exposed
image to the obscured part of the matrix and the continuation
of exposure of the next channel while reading out.

ByuFOSC is very similar to CIGALE (\citealt{1984SPIE..445...37B}).
The instrument provides a useful field of $6\farcm5 \times 13'$
with one pixel equivalent to $0\farcs77$ on the sky.
The field was scanned through 24 steps, the
exposure time was 420~s per channel, providing a total exposure
time of 2~h~48~m. The quality of seeing was about $2''$.
The H$\alpha$ line was isolated with a narrow-band interference
filter centered at 6568.6~\AA{} with a FWHM of 10.3~\AA{}.
General data reduction, night sky lines and stellar continuum subtraction
and velocity field measurements are done similarly to \citet{1987A&A...175..199L}.

\section{Velocity fields and rotation curves}

\begin{figure*}
\begin{center}$
\begin{array}{@{\hspace{0mm}}c@{\hspace{0.0\hsize}}}
\includegraphics[width=0.85\hsize]{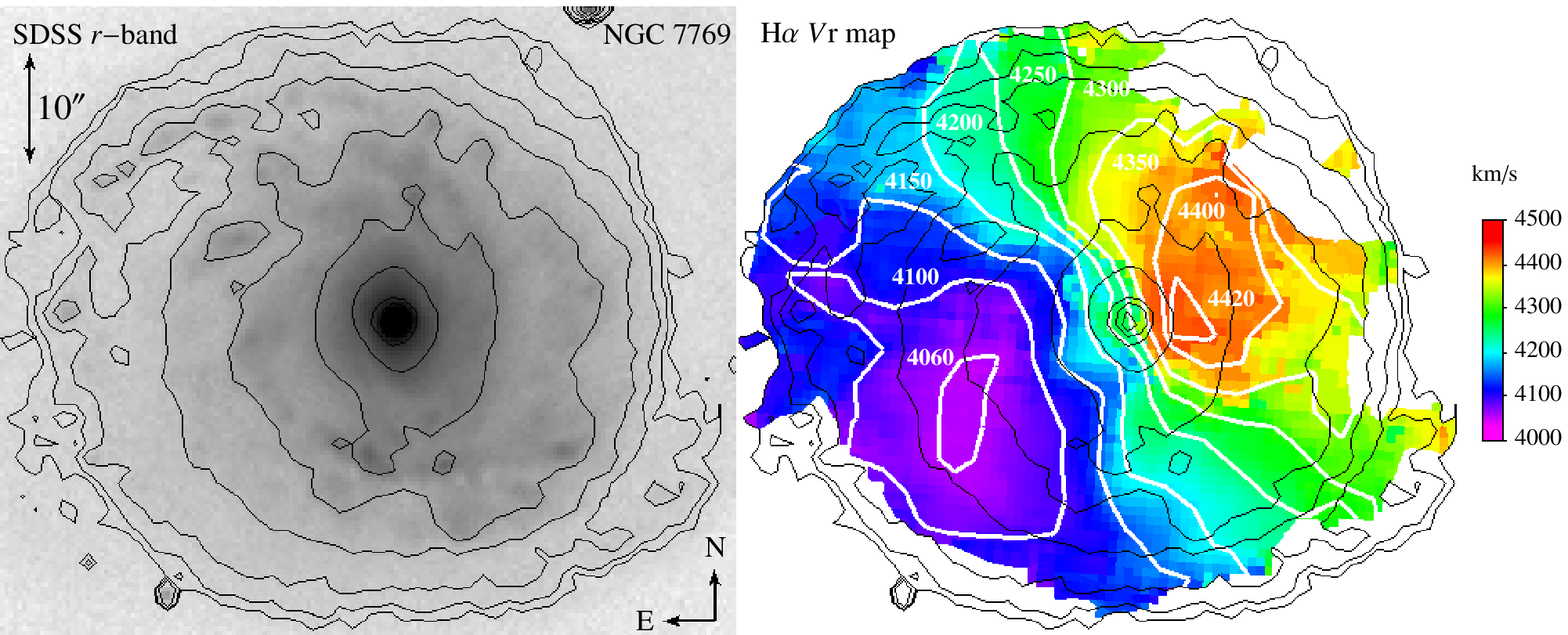}\\
\includegraphics[width=0.85\hsize]{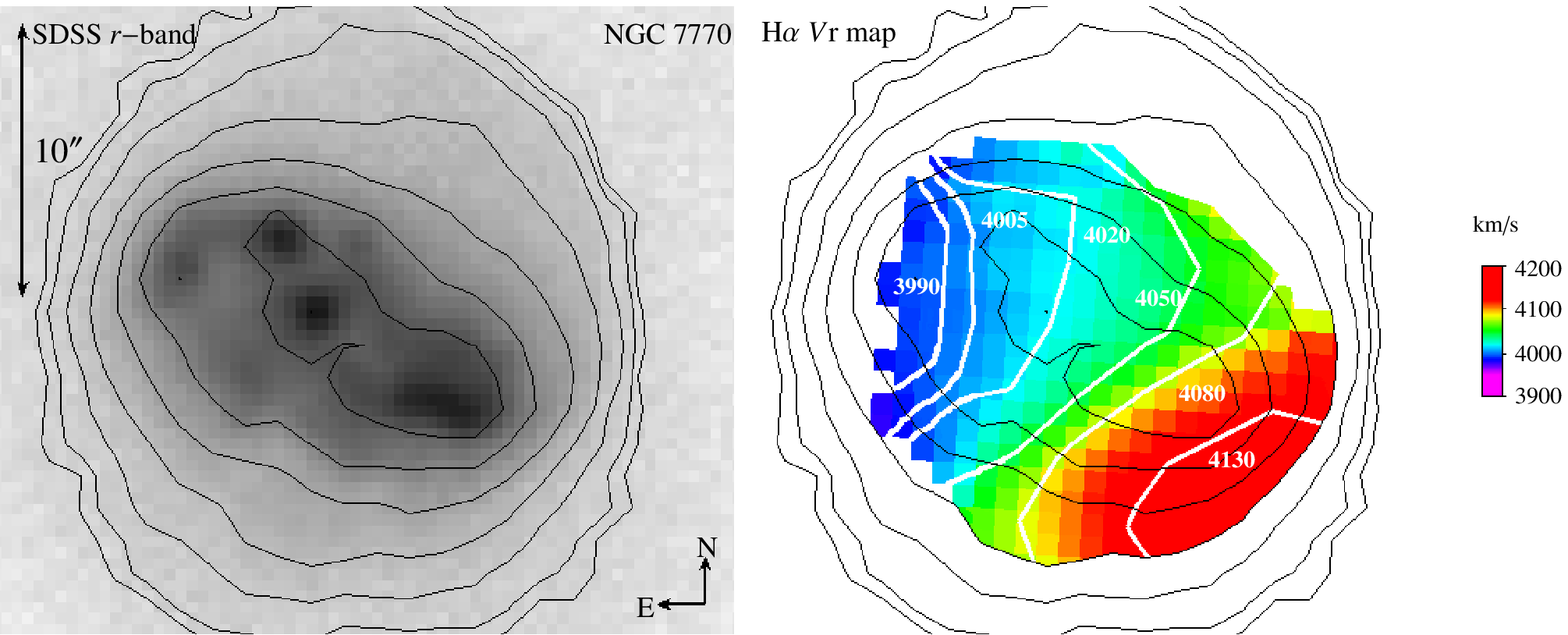}\\
\includegraphics[width=0.85\hsize]{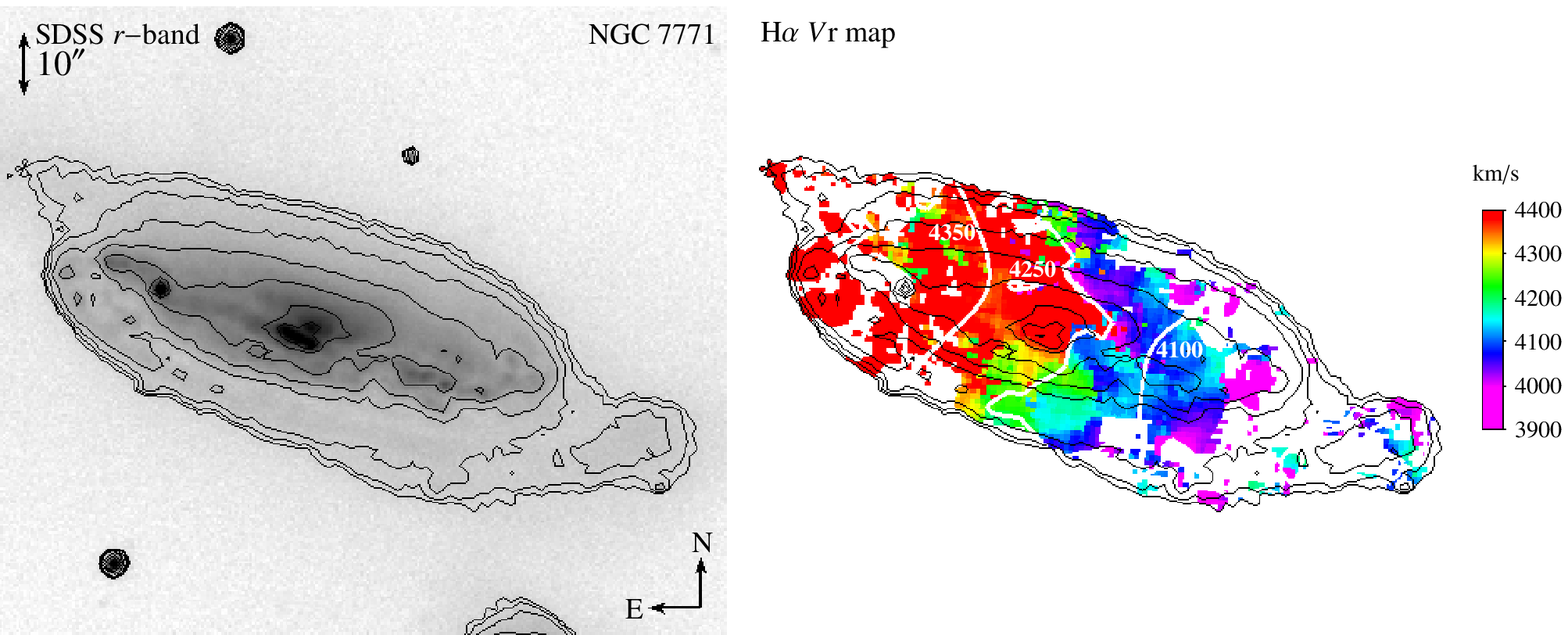}\\
\includegraphics[width=0.85\hsize]{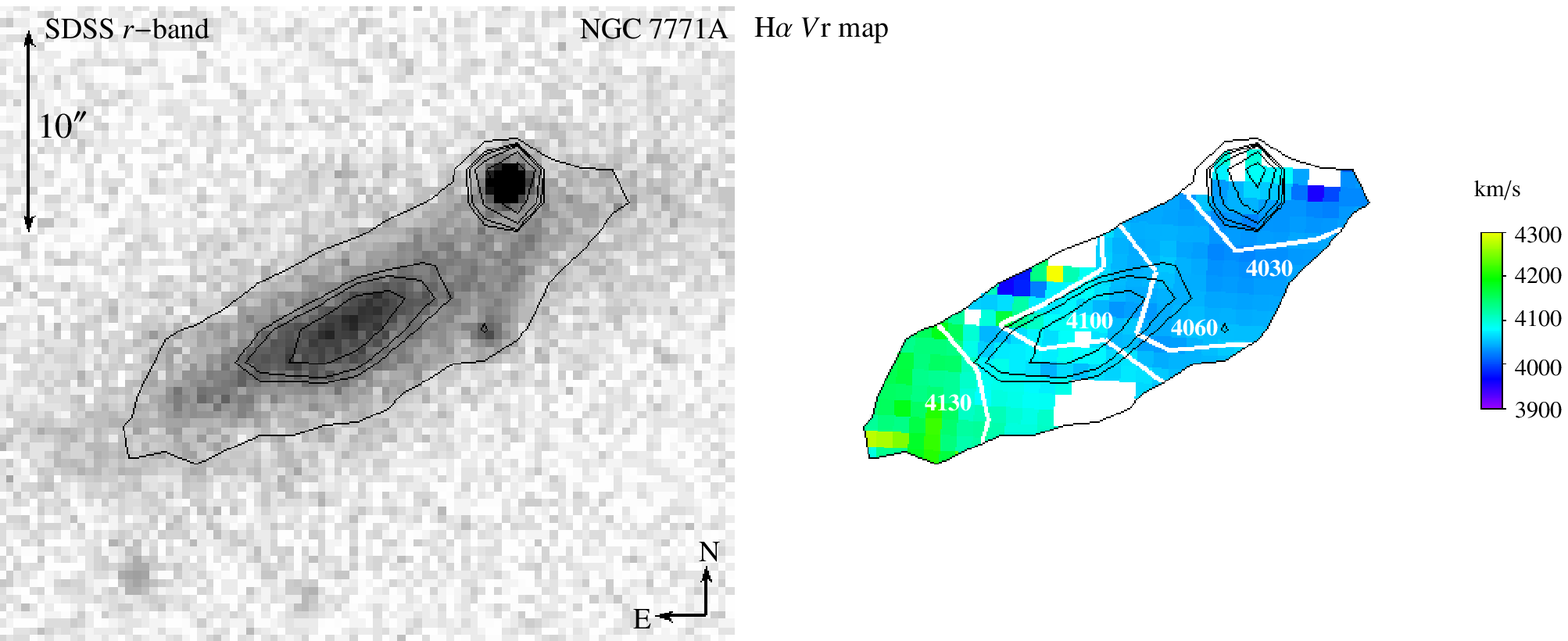}
\end{array}$
\end{center}
\caption{\small{H$\alpha$ velocity fields of galaxies NGC~7769, 7770, 7771 and 7771A,
overlapped by the SDSS \emph{r}-band isophotes (black), and isovelocity contours (white).
The outer isophote corresponds to 22 ${\rm mag~arcsec^{-2}}$ for NGC~7769, 7770 and 7771,
and to 23 ${\rm mag~arcsec^{-2}}$ for NGC~7771A.}}
\label{velr}
\end{figure*}

\begin{figure*}
\begin{center}$
\begin{array}{@{\hspace{0mm}}c@{\hspace{0.0\hsize}}}
\includegraphics[width=0.8\hsize]{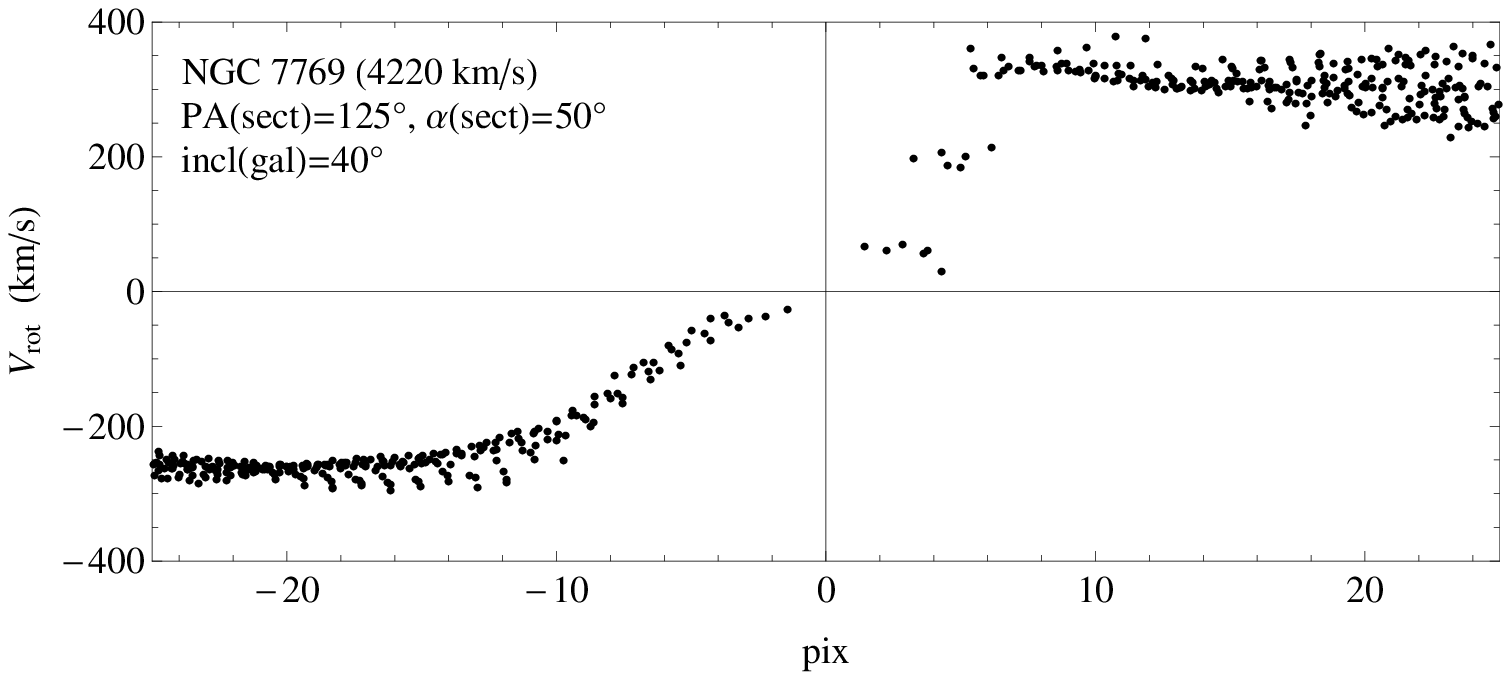}\\
\includegraphics[width=0.8\hsize]{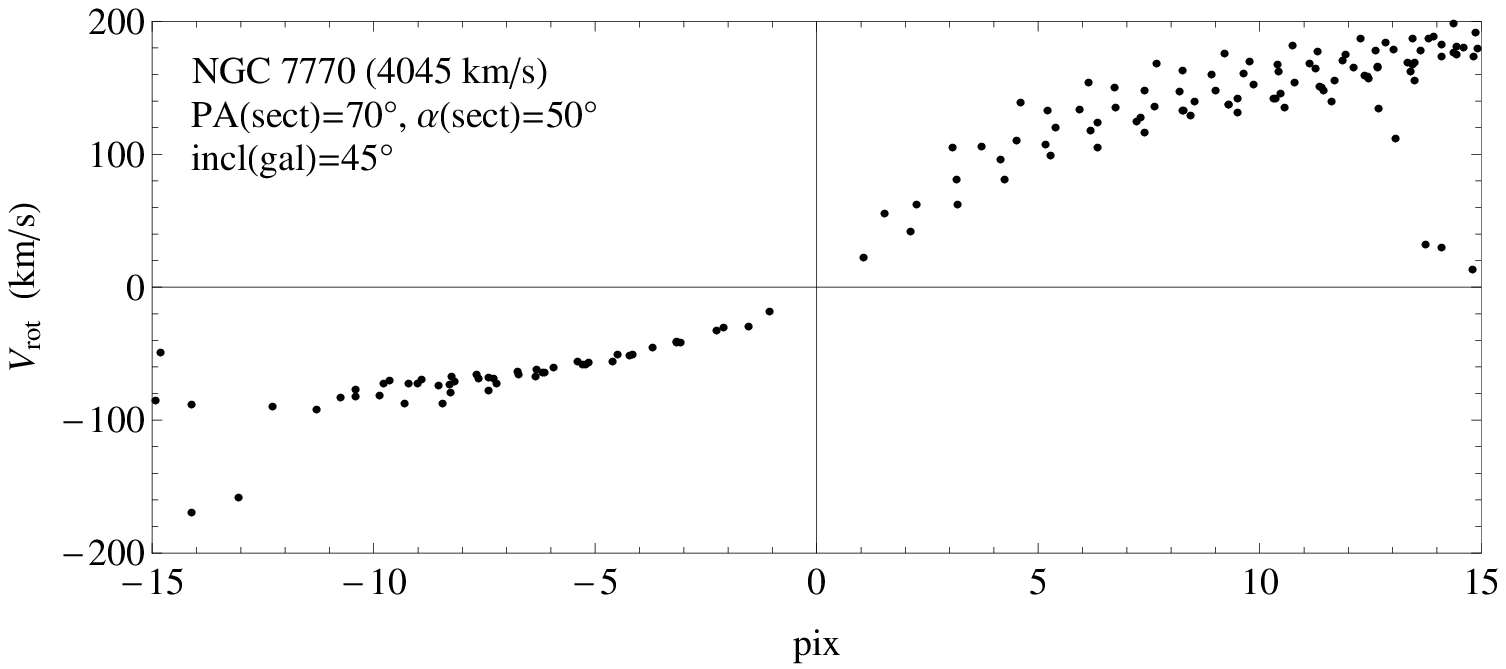}\\
\includegraphics[width=0.8\hsize]{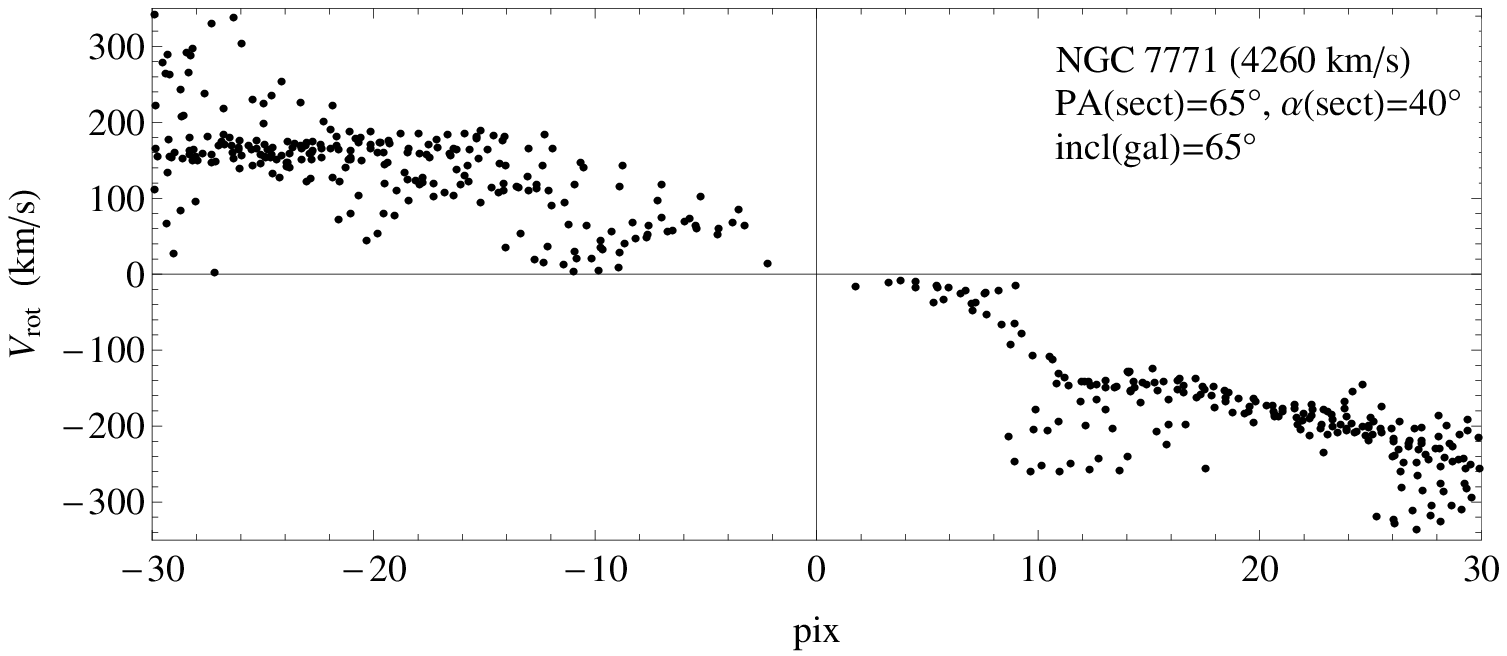}\\
\includegraphics[width=0.8\hsize]{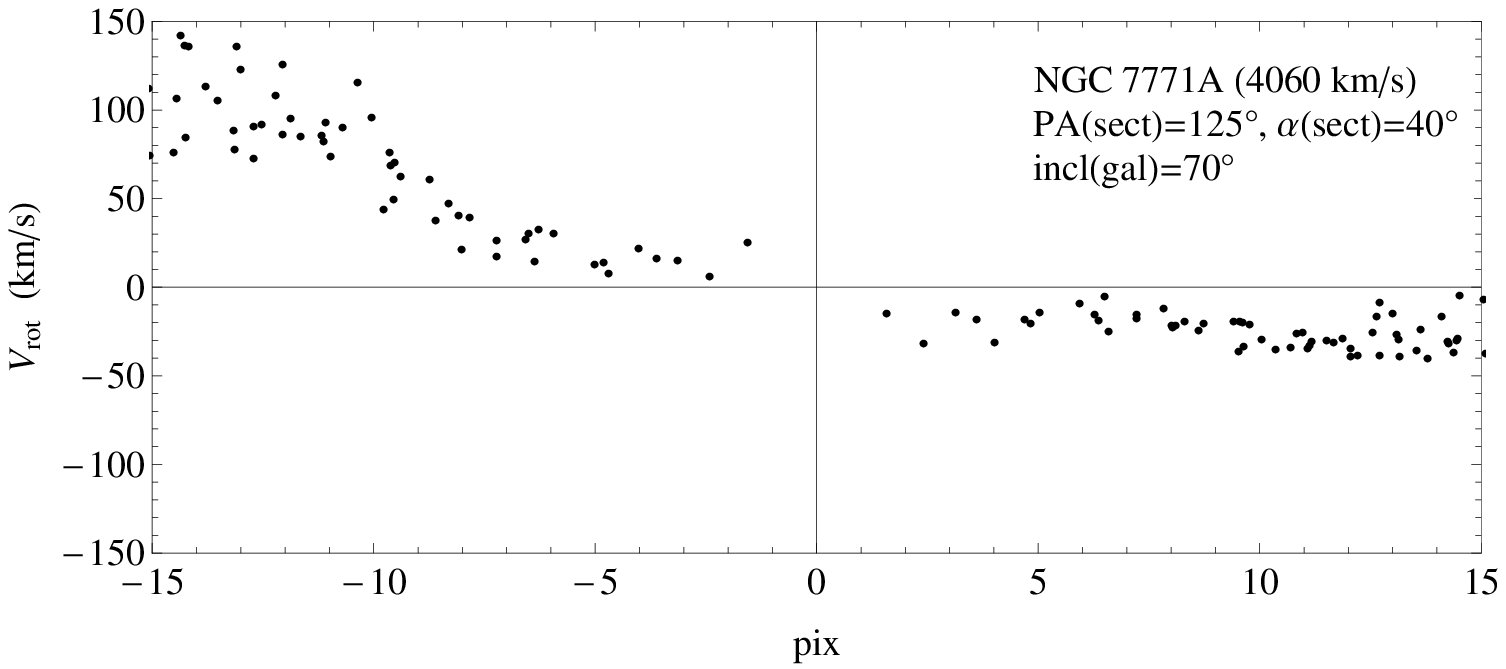}
\end{array}$
\end{center}
\caption{\small{Derived rotational curves of the galaxies. One pixel on the horizontal axis
corresponds to 0.77 arcsec. For each plot, the radial velocity of the galaxy center,
the angle and PA of the sector used to obtain velocity data,
as well as the inclination of galaxy used in the calculations are shown.}}
\label{vrot}
\end{figure*}

Based on the H$\alpha$ velocity fields (Figure~\ref{velr}),
we calculated the rotation curves of the galaxies (Figure~\ref{vrot})
by using data points within sectors along the maximal gradient direction,
see isovelocity contours in Figure~\ref{velr}.


Maximal rotational velocity of NGC~7769 is observed at the radius of
around 15~arcsec from the galaxy nucleus.
The rotational velocities in Figure~\ref{vrot} are in good agreement
with the HI measurements ($316~{\rm km \, s^{-1}}$) in \citet{1993ApJ...419...30C}.
Our measurements of velocities, having a better spatial resolution
compared with those of the previous studies (\citealt{1993ApJ...419...30C,1997AJ....114...77N}),
reveal weak perturbations of the rotation curve of NGC~7769,
which may be caused by interaction with NGC~7771.

The same cannot be said about the velocity field of NGC~7771.
Figures~\ref{velr} and \ref{vrot} show that there are perturbations and
large dispersion in radial velocities at distances
larger than 10-15 arcsec from the nuclei.
This distance is about half radius of the bar.
Evidently, this scatter of radial velocities can be explained
by the fact that part of the arms are included in the sector used
to calculate radial velocities (sector angle is $40^\circ$).
However the asymmetric profile along the major axis suggests that
Northern and Southern arms do not have the same radial velocity profiles.
The asymmetric tidal forces of NGC~7769 and NGC~7770 affecting on NGC~7771,
seem to be a natural cause of that.

The rotation curve of NGC~7770 is significantly skewed.
This is probably because of the strong harassing
interaction with the more massive NGC~7771, see \citet{2012MNRAS.425L..46A}.
The rotation curve of NGC~7771A is typical for a late type Sm galaxy.

By analyzing velocity fields, sizes, and shapes of
spiral arms of NGC~7771 and NGC~7769, in \citet{1997AJ....114...77N},
it has been suggested that NGC~7771 and NGC~7769, which have a 2:1
mass ratio, appear to be having a prograde-retrograde interaction,
with NGC~7769 being the retrograde one.
Our data support this conclusion.
This conclusion is in agreement with the latest models of
galaxy collisions (\citealt{2007A&A...468...61D}) showing that during
direct collisions tidally induced spiral arms are
much longer and brighter than those during retrograde collisions.
We can conclude that galaxies NGC~7769 and NGC~7771 already have passed
the first pericenter stage, however, probably
the second encounter has not yet happened.
The first pericenter distance should have been large enough
(around few sizes of the galaxies), so that large disturbances
in rotation curves have not yet appeared.

\begin{figure}
\centering
\label{profile1}
\includegraphics[width=0.8\hsize]{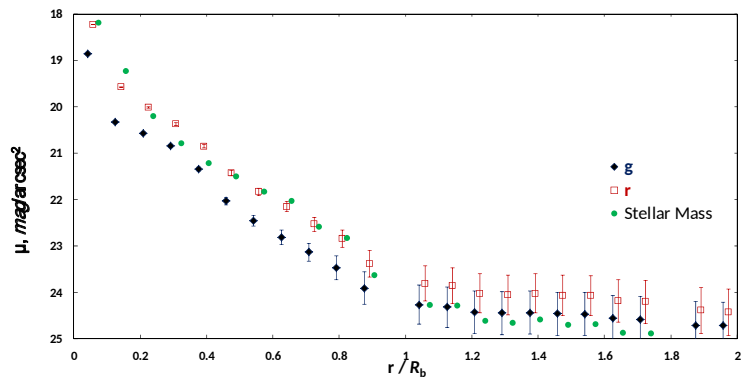}\\
 \parbox{1.0\hsize}{\caption{\small{SDSS \emph{g} and \emph{r} surface brightness profiles and
stellar mass surface density of NGC~7769.
The distance in the horizontal axis is normalized to
the break radius $R_{\rm b} = 0.6~R_{25}$.
Stellar mass surface densities are in arbitrary units
with the same scaling ($-2.5\log [m]$).
Mass density and \emph{r}-band surface brightness
profiles are the same until the break radius,
however, after that distance they are different.}}}
\end{figure}

\section{Photometry and color analysis of NGC~7769}

Studies of color profiles of galaxies allow to find out interesting
features of their evolution, interaction, and star formation
(e.g. \citealt{2008ApJ...683L.103B,2010MNRAS.407..144T}).
NGC~7769 has small inclination, therefore it is possible to
adequately study its radial color profiles, which are quite noteworthy.
In Figure~\ref{profile1}, SDSS \emph{g} and \emph{r}
surface brightness profiles are plotted.
Brightnesses are measured in concentric ellipses having position angle (PA)
of the ellipse with the \emph{g}-band brightness of $25~{\rm mag \, arcsec^{-2}}$.
The characteristic break at the distance $R_{\rm b} = 0.6~R_{25}$
from the nucleus with a smaller gradient after that distance is clearly visible.
However, we should note that the profile of stellar mass surface density,
which is constructed according to the method in \citet{2003ApJS..149..289B}
using the colors and Kroupa initial mass function (IMF), is without a break.
Mass density and \emph{r}-band surface brightness
profiles are the same until the break radius,
however, after that distance they are different.
A Kolmogorov-Smironv test rejects that the brightness profiles of both \emph{r}-
and \emph{g}-bands and mass profile could be from the same parent distribution.

\begin{figure}
\centering
\label{profile2}
\includegraphics[width=0.8\hsize]{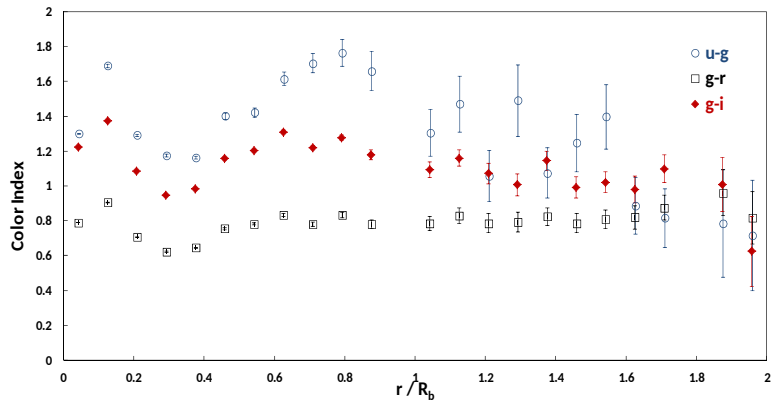}\\
\parbox{1.0\hsize}{\caption{\small{Radial profiles of color indices of NGC~7769.}}}
\end{figure}

In Figure~\ref{profile2}, mean color indices within
the above mentioned elliptical annuli are shown.
At the distance of about $0.3~R_{\rm b}$ (3~kpc) there is a concentric region with minimum color
indices ($g-r=0.62\pm0.02$), which has a higher fraction of young stars.
All the mentioned observational results allow us to classify
NGC~7769 as a galaxy with Type~III profile of surface brightness,
according to classification in \citet{2005ApJ...626L..81E}, see also \citet{2008ApJ...683L.103B}.
Different authors suggested several possible versions
of the origin of the characteristic profile,
all of them implying an external influence on a galaxy.
In case of NGC~7769, we prefer the scenario of \citet{2008ApJ...676L..21Y}
explaining Type~III profiles by a flyby of a neighbor galaxy
with a high-eccentricity orbit.
In our case, it is NGC~7771 galaxy, which has already passed
the phase of first pericenter of encounter with NGC~7769.

\section{Summary}

The quartet of galaxies NGC~7769, 7770, 7771 and 7771A is a system
of interacting galaxies.
In this paper, we present a Fabry-Perot imaging study of the system in
H$\alpha$ line, as well as an analysis
of the color profiles of NGC~7769.
We came to the following main conclusions:

\begin{enumerate}
\item Close interaction between the component galaxies of the system has produced
morphological features that are characteristic of the interactions.
We have detected features such as tidal arms, spiral arms induced by close interaction,
bars and induced star formation.
\item From the results of our interferometric observations, we
obtained the radial velocity profiles of galaxies.
The rotation curve of NGC~7769 is weakly distorted.
The rotation curve of NGC~7771 is strongly
distorted by the tidal arms caused by
direct flyby of NGC~7769 and flyby of a smaller neighbor NGC~7770.
The rotation curve of NGC~7770 is significantly skewed because of
the interaction with the much massive NGC~7771.
\item The radial velocity profiles and morphological disturbances
suggest that the NGC~7769 and NGC~7771
have passed the first pericenter stage,
however, probably the second encounter has not yet happened.
\item The surface brightness profile of NGC~7769 has a characteristic break,
and profiles of color indices have a minimum
at a concentric region of intensive star formation
induced by the interaction with NGC~7771.
\end{enumerate}

Study of such systems with methods combining photometric and visual analysis
is an effective way to clarify features of star formation
in different stages of interaction.
Ongoing and future surveys using integral field spectroscopy will also
allow to explore the spatial distribution of star formation in interacting systems.
\\

\small{
\noindent{{\textbf{Acknowledgements:}}}
\noindent{The authors would like to thank Tigran A. Movsessian for
the help with the observations and data reduction.
They are grateful to the referees for their constructive
comments.
Funding for SDSS-III has been provided by the Alfred P.~Sloan Foundation, the Participating Institutions,
the National Science Foundation, and the US Department of Energy Office of Science.
The SDSS-III web site is \texttt{http://www.sdss3.org/}.}
}

\label{lastpage}
\end{document}